# Programmable Unitary Operations for Orbital Angular Momentum Encoded States

**Shikang Li[1], Xue Feng[1, *], Kaiyu Cui[1], Fang Liu [1], Wei Zhang[1], and Yidong Huang[1]**

[1] Department of Electronic Engineering, Tsinghua University, Beijing, China
Email: * x-feng@tsinghua.edu.cn

**Abstract:**
We have proposed and demonstrated a scalable and efficient scheme for programmable unitary operations in orbital angular momentum (OAM) domain. Based on matrix decomposition into diagonal and Fourier factors, arbitrary matrix operators can be implemented only by diagonal matrices alternately acting on orbital angular momentum domain and azimuthal angle domain, which are linked by Fourier transform. With numerical simulations, unitary matrices with dimensionality of 3×3 are designed and discussed for OAM domain. Meanwhile, the parallelism of our proposed scheme is also presented with two 3×3 matrices. Furthermore, as an alternative to verify our proposal, proof of principle experiments have been performed on path domain with the same matrix decomposition method, in which an average fidelity of 0.97 is evaluated through 80 experimental results with dimensionality of 3×3.

Keywords: unitary transformation, orbital angular momentum, spatial mode, spatial light modulator

## 1. Introduction

Unitary operators are powerful tools in quantum information processing for both investigating non-classical phenomena and exploring quantum computational resources[1]. Great efforts have been made to apply universal linear operation protocols on high-dimensional optical bases, for extending information capacity and enhancing computing versatility[2–4]. Among them, the orbital angular momentum (OAM)[5] modes provide abundant high-dimensional quantum entanglement resources[6–8], while the spiral phase distributions of OAM states have been employed for quantum imaging [9] and remote sensing [10] techniques.

It has been known that any lossless optical setup can be described by a unitary operator[11]. Thus, generally speaking, the generating [12,13], multiplexing [14–16], sorting [17–19] and manipulating [20,21] OAM modes can be considered as certain unitary transformations. Here, we are focused on the arbitrary unitary operator within the OAM-domain. It means that both the input and output states are encoded on a series of optical OAM modes, namely the basis. In our scenario, the weight coefficients of input state would be transformed to those of the output with a unitary operator while the basis keeps. Such a setup or device to perform the unitary operation is quite crucial, since it is the fundamental tool to process the high-dimensional information encoded on OAM modes. However, as analyzed in [21], although there have been great achievements about generating and measuring the OAM modes, how to design an on-demand lossless implementation for an arbitrary unitary operator within OAM-domain has not been well-solved yet. The requirement of closed operation has made it tough to demonstrate unitary operations within OAM-domain, especially with the demand for near-unity success probability at the same time. With the technique of OAM parity sorter[17,18], the four-dimensional Pauli $\hat{X}$ gate and $\hat{Z}$ gate have been reported[20]. However, it would be complicated to construct arbitrary unitary operators based on $\hat{X}$ gates and $\hat{Z}$ gates. Recently, with multi-plane light conversion (MPLC)[21,22] method, cyclic and Fourier transformations are reported with dimensionality up to five. Though a broad variety of unitary operations can be performed with MPLC approach, the transformation efficiency would decrease significantly when the number of OAM modes exceeds four according to the reports[21], thus near-unity efficiency have not been achieved for higher dimensionalities. For techniques with spin-orbital coupling[23,24], only two OAM modes can be utilized due to the two-dimensional encoding space of spin angular momentum so that such scheme is limited and not scalable. Thus, to achieve an OAM-encoded unitary operation scheme with universality, scalability, and near-unity success probability simultaneously is still an open question.

In this work, inspired by the Fourier-diagonal decomposition theorem[25,26] and the demonstration of frequency-domain unitary operation protocols[27,28], we extend this technique to optical encoding basis of OAM domain. A scalable and intrinsic near-lossless scheme for programmable unitary gates based on matrix decomposition[25] is proposed and demonstrated with dimensionality of $3 \times 3$. Moreover, the parallelism of our proposed scheme is also presented with two 3×3 matrices. Limited by the experimental conditions, only numerical simulations are presented for arbitrary OAM-encoded unitary matrices. It should be mentioned that our proposed scheme could also be applied on other two photonic encoding bases that belong to a Fourier transform pair. Thus, as an alternative to verify our proposal, proof of principle experiments have been performed on path domain, in which average fidelity value of $0.97 \pm 0.02$ and transformation efficiency of $0.96 \pm 0.03$ are evaluated through 80 experimental results with matrix dimensionality of $3 \times 3$.

## 2. Theory

Without loss of generality, our target is to implement an arbitrary unitary gate $U$ acting on optical states, which could be expressed as,

$$|\beta\rangle = U|\alpha\rangle \quad (1)$$

where $|\alpha\rangle$ and $|\beta\rangle$ denote the input and output states. Usually the $|\alpha\rangle$ and $|\beta\rangle$ are encoded on a series of optical modes, namely the basis, which could be any degree of freedom, including optical path, frequency, temporal-mode, and transverse-spatial-mode, *et.al*. It should be mentioned that here, the bases of both input and output states are considered as the same. Thus, with a unitary operator of $U$, the weight coefficients of $|\alpha\rangle$ would be transformed to those of $|\beta\rangle$ while the basis is not changed. For most practical applications, $U$ is a finite-dimensional operator and can be mathematically expressed as a unitary matrix with respect to the utilized encoding basis. It has been presented[25] that an arbitrary matrix of dimensionality $N$ can be mathematically decomposed into the production of diagonal matrices and cyclic matrices alternately, where the total number of matrices is $M$, or equivalently saying, $M$ diagonal matrices $D$ spaced by discrete Fourier transformation (DFT) matrix $F$ and its inverse matrix, which is also its Hermitian conjugate $F^\dagger$:

$$\begin{aligned}U_{N\times N} &= FD_M F^\dagger \cdots FD_5 F^\dagger D_4 FD_3 F^\dagger D_2 FD_1 F^\dagger \\ &= (FD_M F^\dagger)\cdots(FD_5 F^\dagger)D_4(FD_3 F^\dagger)D_2(FD_1 F^\dagger)\end{aligned} \quad (2)$$

where $M \leq 2N-1$ [26] denotes the number of diagonal matrices required for the decomposition and the actual value would vary according to the target matrices. Specifically, to establish a unitary operator $U$, the diagonal matrices $D$ in Eq. (2) would also be unitary. To clearly present our idea, the right-hand term in the first line of Eq. (2) is modified as the form of the second one. Particularly, the terms in bracket are diagonal matrices sandwiched by DFT and inverse DFT matrices. Actually, they are also diagonal matrices but acting on a transformed domain, which is determined by the DFT. Thus, Eq. (2) could be understood as that a unitary operator can be decomposed into a series of diagonal matrices in two domains, which are linked with Fourier transformation. A diagonal matrix, which is also referred as a spectral shaper, would not lead to cross-mode effects. Actually, the required cross-components for unitary operations are achieved by DFT matrices in Eq. (2). Thus, if two photonic degree of freedoms (DOFs) are Fourier transform pair, any unitary matrix can be implemented by performing diagonal matrix operations on these two DOFs in succession for the investigated photonic qudits.

It should be mentioned that for the Fourier terms in Eq. (2), it is not required to actually construct the corresponding matrix operation. These Fourier and inverse Fourier operators are naturally inserted in Eq. (2) when the spectral shaping are alternatively performed on two DOFs linked by Fourier transform. Actually, there are some Fourier transform pairs, which have been well investigated in photonic domain, including linear basis ($x$-basis) and transverse wave vector basis ($k$-basis) [29,30], OAM and azimuthal angle [31], and frequency and time-bin, etc. Specifically, arbitrary unitary operators with OAM encoded states can be implemented with diagonal matrices in the OAM and azimuthal angle domain.

With well-defined OAM basis, an arbitrary state vector $|\alpha\rangle$ can be expanded as,

$$|\alpha\rangle = \sum_l c_l |l\rangle \quad (3)$$

where $\{c_l\}$ represent OAM spectrum coefficients and $|l\rangle$ denotes the $l$th eigenstate with OAM of $l\hbar$ carried per photon. As mentioned above, in our scenario, the spectrum coefficients of $\{c_l\}$ would be transformed with a unitary operator of $U$ while the basis of $|l\rangle$ keeps. To perform the decomposition of Eq.(2), both the OAM domain and corresponding Fourier transformation should be employed. It has been investigated that transverse OAM basis $|l\rangle\langle l|$ and angle basis $|\varphi\rangle\langle\varphi|$ are connected via Fourier transformation [31]:

$$\begin{aligned}|l\rangle &= \tfrac{1}{\sqrt{2\pi}} \int_{-\pi}^{\pi} \exp(-il\varphi)|\varphi\rangle d\varphi \\ |\varphi\rangle &= \tfrac{1}{\sqrt{2\pi}} \sum_{l=-\infty}^{+\infty} \exp(il\varphi)|l\rangle\end{aligned} \quad (4)$$

Thus, it is feasible to implement well-designed diagonal matrices within the OAM-domain and angle-domain alternately to achieve any target OAM-domain matrix operator with decomposition form shown in Eq. (2). The question is how to implement the diagonal matrices for both OAM and angle domains. Figure. 1 shows an example of OAM-domain unitary operator with the three-layer scheme corresponding to the case of $M = 3$ in Eq. (2). There are two layers to implement diagonal matrices in angle domain and one layer for OAM domain. As shown later, this scheme could perform most unitary operations with dimensionality of $3 \times 3$, and some particular operations of $4 \times 4$ and $6 \times 6$.

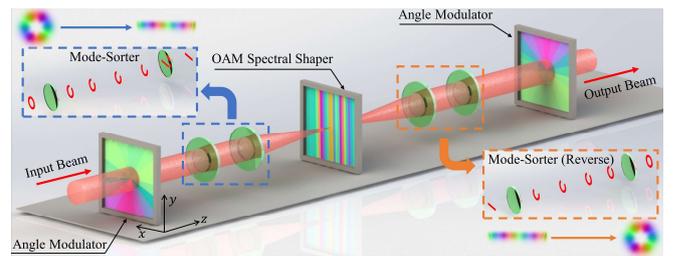

**Figure. 1.** The scheme of OAM-domain unitary operation with three-layer.

In the angle-domain, a diagonal matrix can be readily achieved by a single wave front modulation element such as a spatial light modulator (SLM) [32] or metasurfaces [33]. Here, the SLM is considered and noted as angle modulator in Fig. 1. A typical angle modulation function $f(\varphi)$ can be achieved by the pattern $\Phi_{ANG}(r,\varphi)$ settled on SLM:

$$\Phi_{ANG}(r,\varphi) = f(\varphi), \quad (5)$$

where $r$ and $\varphi$ indicate the radius and azimuthal angle under polar coordinates in the transverse plane according to beam propagation direction, respectively. Thus, the only question is how to realize the required diagonal matrices in OAM-domain, which is also referred as an OAM spectral shaper[34].

Here, a straight-forward method is considered to achieve a programmable and arbitrary OAM spectral shaper. With the help of OAM mode-sorter based on Cartesian to log-polar coordinate transformation[35], each OAM eigenstate can be mapped to different position in the focal plane. As shown in Fig. 1, two static optical elements $\Phi_1(x,y)$ and $\Phi_2(u,v)$ are employed to perform the mode sorter[35,36]:

$$\Phi_1(x,y) = \frac{2\pi a}{\lambda f}\left[y\arctan(\frac{y}{x}) - x\ln(\frac{\sqrt{x^2+y^2}}{b}) + x\right]$$

$$\Phi_2(u,v) = -\frac{2\pi ab}{\lambda f}\exp(-\frac{u}{a})\cos(\frac{v}{a})$$

(6)

where $a$ and $b$ are adjustable parameters, which can be designed according to the beam width of the exact transverse optical field. After passing through these two static optical elements together with an auxiliary lens of focal length $f$, the $l$th OAM eigenstate is sorted to a single spot, in which the center coordinate $x$ in the focal plane reads,

$$C_x(l) = \frac{f\lambda}{2\pi a}l$$

(7)

while the center coordinate $y$ is only determined by parameter $b$ in Eq. (6) and irrelevant to $l$. According to Eq. (7), different OAM modes are spatially separated in the focal plane of the auxiliary lens after the mode sorter consists of $\Phi_1$ and $\Phi_2$ shown in Eq. (6). The coordinate $x$ of the focus center of the $l$th OAM mode is proportional to the value of $l$. As is shown in Fig. 1, an SLM is employed and spatially divided into several blocks along the $x$-axis, which are aligned with the coordinates of the focus centers of each OAM modes. Different phase modulations are settled on these blocks to achieve spectral shaping on OAM modes. In practical implementations, the OAM spectral shaper is determined by the target unitary matrix of $U$. Suppose that the desired spectral shaping function is noted as $g(l)$, the OAM spectral shaper can be implemented with the following wave front modulation function:

$$\Phi_{OAM}(x,y) = g(l), l=\left[\frac{2\pi ax}{f\lambda}\right]$$

(8)

where [.] rounds the value to the nearest integer. As indicated in Fig. 1, such shape function of $\Phi_{OAM}(x,y)$ is achieved with another SLM, which is placed at the focal plane of the auxiliary lens after the second optical element of $\Phi_2$. After OAM spectral shaping, another two static optical elements are required to recover the OAM modes, i.e. to transform the Cartesian coordinate back to log-polar coordinate. The mode-sorter has been proved to be reversible[37] and the static optical elements to perform the reverse mode-sorting are exactly the same as those shown in Eq. (6). Since the static elements for OAM mode-sorter can be replaced by customized spherical lens and cylindrical lens[36], only one programmable SLM performing $\Phi_{OAM}(x,y)$ shown in Eq. (8) would be enough for any OAM spectral shaping task.

## 3. Simulation

We have mathematically modeled the scheme shown in Fig. 1 and simulated the field evolution according to Huygens-Fresnel principle under paraxial approximation. According to the parameters of SLMs (Holoeye Pluto), the simulation area is set to $1080 \times 1080$ pixels with a pixel size of $8 \times 8\mu m^2$. The operation wavelength is considered as 1550nm.

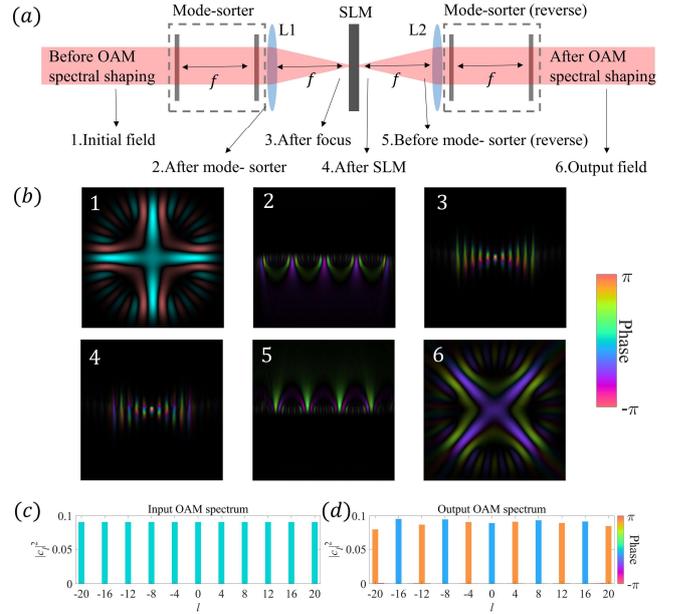

**Figure. 2.** Simulations of OAM spectral shaper with Huygens-Fresnel principle. (a) The scheme of OAM spectral shaper. (b) Step-by-step field evolution of an initial superposed OAM state passing through the spectral shaper. (c) Initial OAM spectrum. (d) Output OAM spectrum. The brightness and color indicate the amplitude and phase distributions, respectively

Firstly, the performance of our proposed OAM spectral shaper is evaluated. As shown in Fig. 2(a), L1 and L2 are lenses with focal length of $f$. A superposed state with 11 OAM modes is shown in Fig. 2(c) as the input state. The mode interval between adjacent OAM encoding channels is chosen as 4 to compress the mode crossing induced by mode-sorter[38]. From left to right, Fig. 2(b) shows field evolution of an initial superposed OAM state passing through the spectral shaper step-by-step (from index 1 to index 6), while the corresponding positions are marked in Fig. 2(a). Here, brightness and color indicate the amplitude and phase distributions, respectively. After the OAM mode-sorter, the initial angular momentum state is mapped to linear state. With the help of lens L1, linear momentum is transformed to linear coordinate in the focal plane. Then the OAM spectral shaping is performed by the SLM settled on the focal plane. After that, the optical vortex is rolled back again by another reversely operating mode-sorter.

The OAM spectral shaper under test is expressed as $g(l) = \pi l/4$. Since only lossless scheme and unitary matrices are concerned, the required diagonal operators in matrix decomposition in Eq. (2) are phase-only and thus referred as $D(g(l)) = Diag(\exp(ig(l)))$ for simplicity. In the following sections, $f(\varphi)$ and $g(l)$ are adopted instead of $\exp(if(\varphi))$ and $\exp(ig(l))$ for angular phase modulation and OAM phase spectral shaping, respectively.

In the case shown in Fig. 2, $g(l) = \pi l/4$ means 0 and $\pi$ phase modulation for adjacent OAM channels. Actually, it is the most challenging case since modulation function is "sharpest" as well as the phase crosstalk is maximum. Such extreme example is considered to examine the crosstalk tolerance of our proposed OAM spectral shaper. Besides, $g(l) = \pi l/4$ is a typical clock matrix and could also be achieved by field rotation of $\pi/4$ rad with a Dove prism[39]. Thus, the OAM spectral shaping effect can be observed intuitively through the rotation between the output and initial field in Fig. 2(b). We also provided the OAM spectrum after spectral shaper in Fig. 2(d). The complex OAM spectrum is calculated with mode-matching method[40]. With this method, the eigenmode expansions of the input and output OAM states during simulation can be evaluated and are expressed as $|\alpha_{in}\rangle = \sum c_{l,in}|l\rangle$ and $|\alpha_{out}\rangle = \sum c_{l,out}|l\rangle$, where $\{c_{l,in}\}$ and $\{c_{l,out}\}$ are input and output OAM spectrum coefficients, respectively. Thus, the spectral shaping function $g(l)$ can be deduced with $g(l) = c_{l,out}/c_{l,in}$. From these simulations, the fidelity value of output OAM spectrum compared with the target one is 0.988, which is estimated according to Eq. (9). It can be seen that the OAM spectral shaper is valid for high-dimensional input states and the unitary operations could be achieved in succession.

Secondly, the OAM spectral shaping function $g(l)$ and two required angular modulation functions of $f_1(\varphi)$ and $f_2(\varphi)$ for a given target matrix are calculated through an optimization algorithm [20]. Our target is to achieve the maximum success probability. The success probability is proportional to the trace value of $VV^\dagger$ subject to $fidelity(V, U) \geq 0.999$, where $U$ and $V$ denote target unitary matrix and designed matrix, respectively. The success probability of implemented matrix after optimization is calculated by $Tr(VV^\dagger)/N$. The fidelity is defined as[41]:

$$fidelity(U,V) = \left| \frac{Tr(U^\dagger V)}{\sqrt{Tr(U^\dagger U) \cdot Tr(V^\dagger V)}} \right|^2 \quad (9)$$

The designed matrix $V$ in Eq. (9) is calculated by OAM spectral shaping function $g(l)$ and two angular modulation functions $f_1(\varphi)$ and $f_2(\varphi)$. Here, we give an example with dimensionality of $2 \times 2$:

$$\begin{pmatrix} \ddots & & & \\ & V_{2\times2}^{11} & V_{2\times2}^{12} & \\ & V_{2\times2}^{21} & V_{2\times2}^{22} & \\ & & & \ddots \end{pmatrix}_{K \times K} = F^\dagger D(f_1(\varphi)) F \times D(g(l)) \times F^\dagger D(f_2(\varphi)) F \quad (10)$$

The number $K$ in Eq. (10) is the dimensionality of numerical optimization. The central $2 \times 2$ region in the left side of Eq. (10) is the optimized matrix $V$. We note that $K = N + 2d$, where $N$ is the dimensionality of target matrix ($N = 2$ for the case in Eq. (10)) and $d$ is the number of guard bands. The guard bands are required to eliminate cut-off effects, which arises from the differences between the finite-dimension discrete Fourier transform required in Eq. (2) and the infinite-dimension nature of the Fourier relation of OAM eigenstates and angle eigenstates expressed in Eq. (4). In this work, both for the OAM-encoding and path-encoding, the number $K$ is chosen to be $K = 64$ during the numerical calculation of spectral shaping functions. The DFT matrices in Eq. (10) are normalized with $1/\sqrt{K}$ in amplitude so that $Tr(VV^\dagger)/N \leq 1$. Besides, we have $Tr(VV^\dagger)/N = 1$ if and only if all other elements that share the same rows or columns with the central $N \times N$ block are zero.

The angular modulation functions $f_1(\varphi)$ and $f_2(\varphi)$ are quasi-continuous since the $[0,2\pi)$ interval is divided into $K$ bins. The angular modulation functions are represented by sine expansion:

$$f(\varphi) = \sum_{n=1}^{p} A_n \sin(n\varphi + \theta_n) \quad (11)$$

The parameters $\{A_n\} \in [0, m\pi)$ and $\{\theta_n\} \in [0, 2\pi)$ are to be determined. The scalar $m$ may grow as the size $N$ of target matrices increases, to assure enough interactions between farther OAM channels. Here $m = 1$ and $m = 2$ are chosen for 2-dimensional and 3-dimensional unitary matrices, respectively. By choosing finite series expansion of sine expression, a relatively smooth modulation function can be obtained. Thus, the corresponding phase masks could have lower spatial resolution according to the Nyquist criterion and can be realized with more facility. The number $p$ in Eq. (11) is determined by the dimension $N$ of target matrix since higher orders of sine components in angular modulation functions allow interactions between farther OAM eigenstates. $p = 3$ is chosen in our simulations. The entire optimization is done with the MATLAB "fmincon" function. To further explain the optimization procedure and provide an executable example, the functions $g(l)$, $f_1(\varphi)$ and $f_2(\varphi)$ required for $2 \times 2$ and $3 \times 3$ Hadamard matrices $H_2$ and $H_3$ as well as modulation functions settled on SLMs are shown in Fig. 3.

$$H_2 = \frac{1}{\sqrt{2}}\begin{pmatrix} 1 & 1 \\ 1 & -1 \end{pmatrix}, H_3 = \frac{1}{\sqrt{3}}\begin{pmatrix} 1 & 1 & 1 \\ 1 & e^{i2\pi/3} & e^{i4\pi/3} \\ 1 & e^{i4\pi/3} & e^{i2\pi/3} \end{pmatrix} \quad (12)$$

The schematic setup for OAM-domain matrix transformation is shown in Fig. 3(a). Three SLMs labelled as SLM1, SLM2, and SLM3 perform the first angular modulator, OAM spectral shaper, and the second angular modulator, respectively. The colored phase modulation patterns corresponding to each SLM are also shown. For Hadamard matrix $H_2$, the optimized angular modulation functions are shown under polar coordinates in Fig. 3(b), where the red and blue solid lines indicate $f_1(\varphi)$ and $f_2(\varphi)$, respectively. The unit of polar axis in Fig. 3(b) is rad. The optimized OAM spectral shaper noted as $g(l)$ is shown in Fig. 3(c), where the OAM eigenstates of $l = 0$ and $l = 1$ are chosen as encoding channels for $H_2$. In Fig. 3(c), 10 guard bands are plotted. Since the optimized spectral shaping functions tend to converge for more guard bands, much less OAM channels need to be modulated even when number $K$ in Eq. (10) is chosen to be as large as 64.

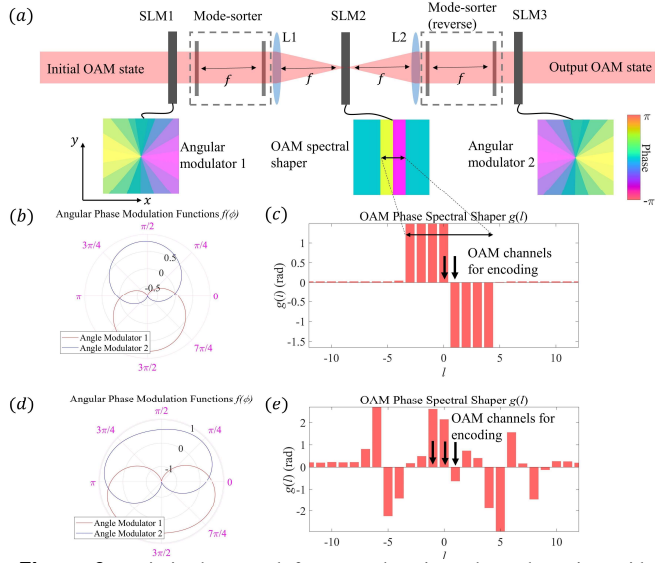

**Figure. 3.** Optimized approach for OAM-domain Hadamard matrices with dimensionality of $2\times2$ and $3\times3$. (a) Modelled setup for achieving OAM-domain matrix transformation. (b) Optimized angular modulation functions $f_1(\varphi)$ and $f_2(\varphi)$ for Hadamard matrix $H_2$. (c) OAM spectral shaper $g(l)$ For $H_2$. (d) and (e) Optimized modulation functions for Hadamard matrix $H_3$.

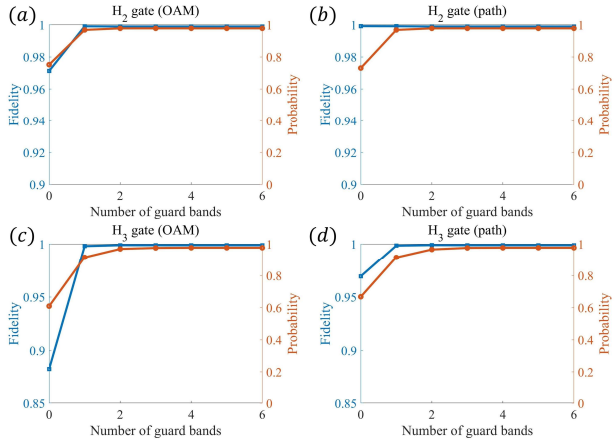

**Figure. 4.** Fidelity and probability of unitary transformation as functions of numbers of guard bands. The target matrices are chosen as $H_2$ and $H_3$ in (a, b) and (c, d), respectively. The channels beyond guard bands are unchanged in (a, c) and filtered out in (b, d).

Numerical estimation is made to quantize the influence of the number of guard bands on transformation fidelity and success probability. For those encoding channels beyond guard bands, there are two situations that have to be considered. The signals that enter those channels are left unchanged for OAM-encoding since those OAM channels always exist objectively. For path-encoding employed in experiments in section 4, the channels beyond guard bands are filtered out since the number of accessible encoding channels are determined by the size of SLMs (Fig. 7 in section 4). For both cases, the fidelity and probability of unitary transformation as functions of numbers of guard bands are plotted in Fig. 4. The target matrices are chosen as $H_2$ and $H_3$ in Fig. 4(a, b) and Fig. 4(c, d), respectively.

According to the numerical results shown in Fig. 4, the condition of $d \geq N-1$ would be enough for most practical cases. Thus, only two or three guard bands are required for practical $H_2$ gate. Due to the value of $K = 64$, the optimized angular modulation function shown in Fig. 3(b) seems quasi-continuous. Those optimization results for Hadamard matrix $H_3$ are shown in Fig. 3(d) and (e), where OAM eigenstates $l = -1$, $l = 0$ and $l = 1$ are chosen as encoding channel. According to the optimization, the fidelity above 0.999 and success probability above 0.970 are achieved simultaneously for $H_2$ and $H_3$. After the phase patterns settled on SLMs are obtained, the field evolution of Hadamard matrix $H_2$ has been calculated according to Huygens-Fresnel principle. The OAM eigenstates of $l = -3$ and $l = 3$ are chosen to be the encoding channels. The transformation fidelity is calculated as 0.998.

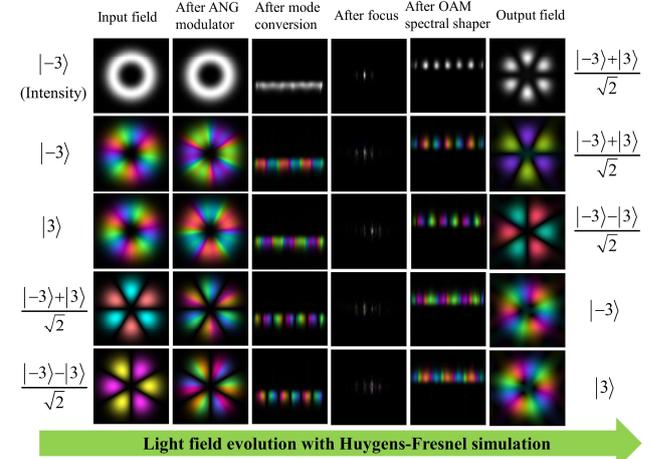

**Figure. 5.** Step-by-step field evolution of OAM-domain Hadamard matrix.

In Fig. 5, each row represents the field evolution of different initial OAM states and six columns are corresponding to the initial field, field after SLM1, field after mode-sorter, field before SLM2, field before L2, and output field (from left to right). All rows are achieved with the same angular phase patterns so that the same OAM phase spectral modulation functions are obtained. Moreover, in each map, the brightness and color are corresponding to the amplitude and phase, respectively. It should be mentioned that there is no need for post-selection to perform OAM-domain closed transformation with our proposal and thus the near-unity success probability is preserved. Thus, our results for implementing OAM-domain Hadamard matrix (or OAM beam splitter) may inspire some new applications such as observation of OAM-domain Hong-Ou-Mandel (HOM) effect[42] and OAM-encoded measurement-device-independent quantum key distribution (MDI-QKD)[43].

Additionally, another important feature of our proposal is parallel computing, which is considered as an advantage of optical information processing. With our scheme, the same matrix could be applied on different OAM channels simultaneously without any extra hardware cost. To further explain this, the optimized OAM spectral shaping function

$g(l)$ is shown in Fig. 6(a) for two $H_2$ gates acting on two groups of OAM channels with $l = -3\&-2$ and $l = 4\&5$. For SLM modulation functions, the only change is to replace the OAM spectral shaping function $g(l)$ by

$$conv(g(l), \delta(l-l_1) + \delta(l-l_2) + \cdots) \qquad (13)$$

where $conv(.)$ denotes convolution and $\delta(l-l_1)$ denotes the Dirac delta function. The OAM charge $l_1, l_2$ ... are the central OAM channels utilized by each identical matrix. It is unnecessary to change the angular modulation functions $f_1(\varphi)$ and $f_2(\varphi)$ for such parallel processing. Two $H_2$ gates acting on OAM channels with $l = -14\&-10$ and $l = 10\&14$ are simulated and the entire $4 \times 4$ matrix operator is evaluated and shown in Fig. 6(b). Furthermore, the simulated results of dual $H_3$ gates are shown in Fig. 6(c). The utilized OAM channels are $l = -18\&-14\&-10$ and $l = 10\&14\&18$. The $H_3$ gate is also noted as $3 \times 3$ DFT. In Fig. 6(c), the $H_3$ gate is tested with input superposed OAM states of $|\omega_1\rangle, |\omega_2\rangle$ and $|\omega_3\rangle$ one-by-one, while $|\omega_n\rangle$ denotes the $nth$ column of $H_3^\dagger$. With this test, the phase accuracy of the unitary matrix can be evaluated. The colored bars and empty bars in Fig. 6(c) indicate input complex OAM spectrum and output OAM spectrum, respectively. It should be mentioned that the simulation in Fig. 6 is made under the same three-layer model as shown in Fig. 3(a). It has been observed that the OAM-domain parallel gates work well without any extra hardware cost.

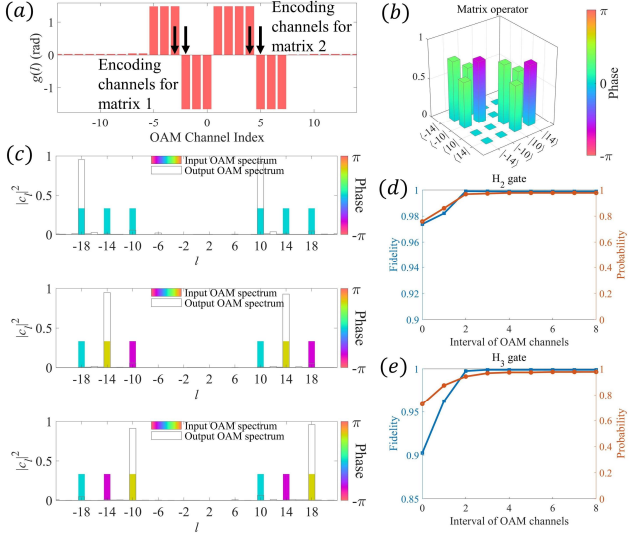

**Figure. 6.** The OAM-domain parallel $H_2$ gates and $H_3$ gates achieved by the three-layer structure. (a) OAM spectral shaping function for dual $H_2$ gates. (b) The entire $4 \times 4$ matrix operator made up of two $H_2$ gates. (c) Dual $H_3$ gates tested with input superposed OAM states of $|\omega_1\rangle, |\omega_2\rangle$ and $|\omega_3\rangle$ one-by-one. (d) and (e) performances of parallel $H_2$ and $H_3$ gates as functions of OAM channels separation, respectively.

In Fig. 6(d) and (e), the performances of parallel $H_2$ and $H_3$ gates are plotted as functions of OAM modes separation, respectively. It could be found that the parallel gates operate independently with negligible crosstalk if the interval of utilized OAM channels is larger than two times of the number of necessary guard bands.

Actually, the number of parallel matrices is only limited by the number of available OAM channels. Here, the mode interval of adjacent OAM channels is settled as 4 to avoid mode crosstalk induced by OAM mode-sorters. The OAM channel resources would be utilized more efficiently if some efforts are made to reduce the crosstalk[44]. According to the mathematic decomposition form in Eq. (2) as well as the numerical optimization form in Eq. (10), it is quite straight forward to extend this three-layer structure in Fig. 1 to larger scales. With alternately cascaded angular spectral shaper layers and OAM spectral shaper layers, the system arrangement would grow with complexity of $(2N - 1)$ [26] according to the dimensionality $N$ of the target unitary matrix.

In our design, the free space focus systems shown in the conceptual setup in Fig. 3(a) only serve for the OAM mode-sorters. Besides this, there is no need for free-space propagation or extra Fourier transformation blocks between two adjacent layers of spectral shapers. For OAM domain unitary operations, this is one of the main differences to the MPLC method[21,22], in which some free-space propagation between adjacent layers is required. It should be mentioned that our proposal is not constrained by the actual method of OAM spectral shaping. In our simulation, the OAM spectral shaping is achieved with OAM mode-sorters, which would induce extra experimental difficulties. Actually, the system arrangement shown in Fig. 1 could be largely simplified with compact OAM spectral shaper. For some specific target matrices, the recently reported single-step OAM shaper[34,45] may be employed.

## 4. Experiment

Due to the lack of OAM mode-sorter elements, which requires advanced 3D printing technology, proof-of-principle experiments in path domain are performed to verify our proposed scheme. To utilize the matrix decomposition in Eq. (2), we notice that there is a one-to-one mapping from transverse wave vector in the object plane to transverse coordinate in the focal plane. To encode qudits, the path basis can be defined by a series of transverse coordinates lying on $y-$axis in the focal plane. Thus, the optical field $\widetilde{E_o}$ in the object plane is linked to the aforementioned path-encoded state vector through the Fourier transformation and expressed as:

$$\widetilde{E_o}(y) = \sum_{n=-\infty}^{+\infty} c_n \exp(-ink_y y) \qquad (14)$$

where $c_n$ denote the complex amplitude of the $n$th path-encoding channel in the focal plane. It should be mentioned that $\widetilde{E_o}$ and the path-encoded state vector of $\sum c_n |n\rangle$ are equivalent expressions of the same state under momentum space and position space, respectively. Therefore, unitary operations can be achieved by alternately performing spectral shaping in momentum space and position space. The distance between adjacent paths is noted as $L$, while the distance between adjacent SLMs is set to $2f$. This leads to a corresponding transverse wave vector of

$$k_y = \frac{L}{2f}\left(\frac{2\pi}{\lambda}\right) = \frac{\pi L}{\lambda f} \qquad (15)$$

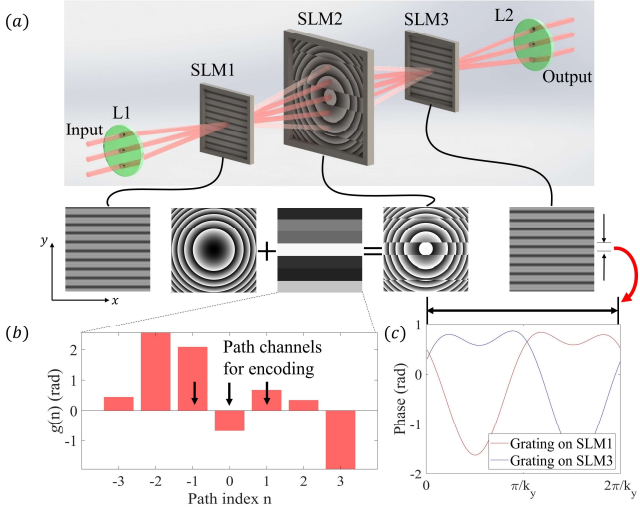

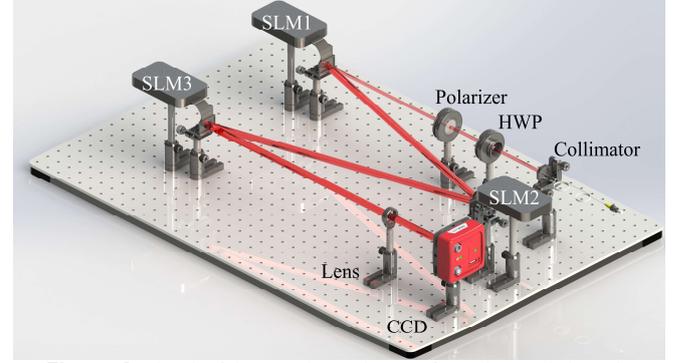

**Figure. 7.** (a) Conceptual set-up of path-encoded matrix transformation, and the modulation functions settled on SLMs. (b) Optimized phase modulation function settled on SLM2 for implementing $H_3$ gate. (c) One period of optimized phase gratings settled on SLM1 and SLM3 for $H_3$ gate.

**Figure. 8.** Sketch of the experimental set-up for path-encoded matrix transformation. HWP: Half Wave Plate, SLM: Spatial Light Modulator. CCD: Charge-Coupled Device camera.

The conceptual setup is shown in Fig. 7(a). The modulation functions on SLM1 and SLM3 are settled as following to introduce the required transverse wave vector components.

$$f(y) = \sum_{n=1}^{p} A_n \sin(n k_y y + \theta_n) \qquad (16)$$

Eq. (16) is very similar to the azimuthal phase modulation function in Eq. (11) for OAM-domain matrix transformation. The parameters $\{A_n\} \in [0, 2\pi]$ and $\{\theta_n\} \in [0, 2\pi]$ are to be determined by optimizations and possess the same meanings as shown in Eq. (11). The only difference is that the azimuthal coordinate $\varphi$ is replaced by linear coordinate $y$. The modulation function on SLM2 is settled as

$$\Phi(x, y) = g\left(\left[\frac{y}{L}\right]\right), \qquad (17)$$

which is also a direct mapping of Eq. (8). The function of $g(n)$ represents the phase modulation applied to the $n$th path channel. Practically, $g(n)$, $f_1(y)$, $f_2(y)$ are optimized with the same method as $g(l)$, $f_1(\varphi)$, $f_2(\varphi)$ in OAM-domain unitary transformation. The $H_3$ gate is shown as an example to clarify it. The optimized $g(n)$ is shown in Fig. 7(b), where path channels n = −1&0&1 are chosen for encoding. The $f_1(y)$ and $f_2(y)$ functions are shown in Fig. 7(c) as phase-only gratings programmed on SLM1 and SLM3, respectively. Gratings can be considered as spectral shaping in transverse momentum space. Here, only one period of grating is plotted for simplicity and clarity. A lens with focal length of $f$ is added on SLM2 to maintain the Fourier relation between optical field on adjacent SLM planes. It should be mentioned that the one-dimensional Fourier relation is employed here, which can be expressed as a Fourier transform matrix and is different from the two-dimensional Fourier transform that is achieved by free space propagation. Two extra lenses noted as L1 and L2 in Fig. 7(a) are employed to compensate for transverse wave vector so that the setup can be scalable. As indicated in Eq. (14), we are only interested in the field variations along $y$ − axis in the transverse plane. The field along $x$ − axis is simply set as Gaussian shape and some ancillary lenses are programmed on SLMs to compensate the natural expansion of Gaussian beams.

80 different unitary transformations have been performed with the experimental setup as shown in Fig. 8, while 75 of them are randomly generated. For each different target matrices, a particular set of modulation functions has to be optimized. The number of guard bands are set as $d = 2$. As shown in Fig. 8, a laser beam of zero-order Gaussian profile operating at 1550 nm (RIO Orion) is incident to the optical system through a collimator. Three SLMs (Holoeye Pluto) operating with reflective mode are utilized and the focal length $f$ mentioned before is set as 40 cm. The distance $L$ between adjacent paths is designed as 2 mm considering the reflective area of SLMs. Specially, the input state is both generated and switched by SLM1. The transverse $k$-space expression $\widetilde{E_o}$ in Eq. (14) of the input state, or equivalently saying, the interference pattern of the input beams at SLM1 plane in conceptual setup in Fig. 7(a) is pre-calculated and directly settled on SLM1 to modulate the incident Gaussian beam from the collimator in experimental setup shown in Fig. 8.

The experimental results are summarized in Fig. 9(a) and (b). The distributions of fidelity and success probability of implemented matrices are shown as solid bars, respectively. Here, the fidelity is evaluated by a charge coupled device (CCD) camera placed at the output port in the experimental setup shown in Fig. 8. Since the target matrix $U$ is unitary, the value of $fidelity(V, U)$ between $U$ and implemented matrix $V$ equals to $fidelity(VU^\dagger, UU^\dagger)$ according to Eq. (9).

$$fidelity(U, V) = fidelity(UU^\dagger, VU^\dagger) = \left|\frac{Tr(VU^\dagger)}{\sqrt{NTr[(VU^\dagger)^\dagger(VU^\dagger)]}}\right|^2 \qquad (18)$$

The amplitude of $VU^\dagger$ can be directly measured by the CCD camera. Actually, to measure $VU^\dagger$ is a phase test. During the test, the input states of $|\psi_1\rangle$, $|\psi_2\rangle$ and $|\psi_3\rangle$ are settled as the columns of target matrix $U^\dagger$. The results would also form an identical matrix if the implemented matrix $V$ is sufficiently accurate, as shown in Fig. 9 (c~h). The phase differences among diagonal elements of $VU^\dagger$ are assumed to be zero during this estimation because subsequent phase shifters could be used for error correction[46]. Besides, according to Eq. (2), the implementing errors of any diagonal matrix $D$ would influence both relative phase within matrix rows and relative phase between matrix rows simultaneously. Thus, if the relative phase within matrix rows is accurate evidently, it is unlikely to remain a large relative phase between matrix rows.

The amplitude of $V$ is scanned with input state $|1\rangle$, $|2\rangle$ and $|3\rangle$, providing an intuitive vision of absolute value of matrix elements. The success probability is calculated by the output optical power from the three path-encoding channels normalized by the total output power received by the CCD camera.

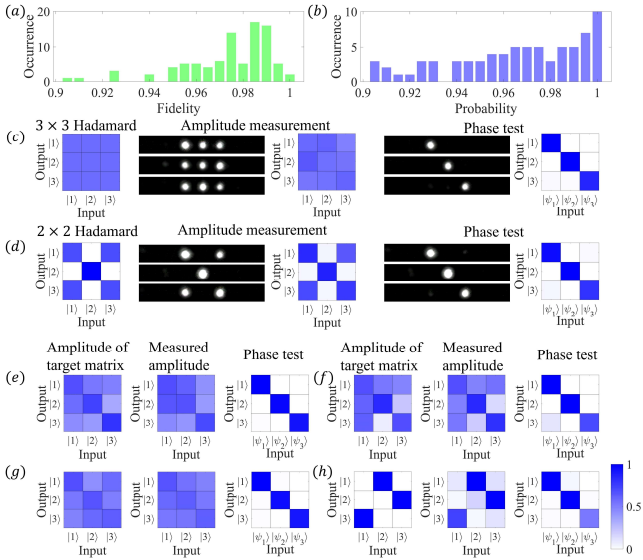

**Figure. 9.** Experimental results of path-encoded matrix transformation. (a) The distributions of (a) fidelity and (b) success probability for 80 different tests. The typical experimental results for (c) Hadamard matrix $H_3$, (d) $H_2$ and (e~h) some other unitary operators. For each group, there the amplitude of target matrix $U$, the measured amplitude and the measured phase of implemented matrix $V$ from left to the right. The color bar is also shown.

The average value of fidelity is $0.97 \pm 0.02$ among 80 different tests. The success probability is shown in Fig. 9(b) with an average value of $0.96 \pm 0.03$. The main reason of deteriorated probability is that part of optical energy entered guard bands and failed to vanish through interference. The experimental results of $H_3$ gate and $H_2$ gate (acting on the first and third channel) are shown in Fig. 9(c) and (d), respectively. The original data captured by CCD camera are also shown. As concrete examples, the data of another four typical unitary matrices are shown in Fig. 9 (e ~ h), while the fidelity value ranges from 0.96 to 0.99.

A brief error analysis for experimental results is as follows. First, the conditioned maximum algorithm is utilized to determine the modulation functions of $g(n)$, $f_1(y)$, $f_2(y)$. The fidelity values are conditioned by $fidelity \geq 0.999$. As a result, the average probability value tends to be lower than the average fidelity value. Second, a five-layer structure is sufficient for arbitrary $3 \times 3$ matrices according to the decomposition theory[26]. Though the three-layer structure employed in this work is enough for most cases, the fidelity and probability could not both achieve 1 at the same time for some particular target matrices, hence there is a tradeoff. Third, experimental errors in terms of optical misalignment and nonlinear effects induced by CCD camera would also deteriorate both the fidelity and probability. The saturated effect of some CCD images will result in lower fidelity and efficiency estimation. The fidelity is calculated through phase test shown in Fig. 9(c) and (d). The saturated effect induces reduction of intensity of main lobs but has little influence on side lobs, thus lowers the weight of main lobs as well as the value of $Tr(VU^\dagger)$. The detailed evaluation considering a $2 \times 2$ case is below.

After normalization, suppose that an averaged reduction of $E_{reduct}$ is induced for all diagonal elements of $VU^\dagger$ due to saturation. At the same time, the difference among diagonal elements of $VU^\dagger$ would decrease by $E_{even}$. The $E_{even}$ value is comparable with $E_{reduct}$ value and we have $E_{even} \leq E_{reduct}$ if the derivative value of the saturation function is less than 1. According to the definition of fidelity in Eq. (18), the numerator term of $Tr(VU^\dagger)$ would decrease by $2E_{reduct}$, while the denominator term of $Tr[(VU^\dagger)^\dagger VU^\dagger]$ would decrease from about $(1 + E_{even})^2 + (1 - E_{even})^2$ to $2$, assuming that the values of diagonal elements of $Tr(VU^\dagger)$ are about $1$. The denominator term would decrease by $2E_{even}^2$, which is one order smaller than the reduction of numerator. The square root calculation in denominator further shrinks the reduction. Hence, the saturated effect will result in lower fidelity estimation.

The amplitude measurement shown in Fig. 9(c) and (d) are used to evaluate transformation efficiency (success probability). The saturated effect lowers the measured optical power from the three path-encoding channels and decreases the efficiency value.

It could be concluded that the three-layer structure is valid to achieve both high fidelity and high success probability for most 3-dimensional unitary matrices. Thus, the experimental results of path-encoded unitary transformation are important evidences to prove the versatility of our OAM domain proposal.

## 5. Discussions

First, the relations and differences between our work and the frequency-domain protocols are discussed. Our method shares the same basic mathematical theorem[25] of matrix decomposition with Lougovski's work[47]. In their work, the frequency-time Fourier relation is employed. We notice that frequency encoding is different from other photonic encoding since energy conservation is not satisfied for unitary linear operations such as the Pauli $\hat{X}$ gate. Thus, it cannot be achieved by only passive setup. In this work, we focus on OAM and path domains employing the OAM-angle Fourier relation and the position-momentum Fourier relationship. Correspondingly, the unitary operations within both OAM and path domains could be implemented with passive optical elements.

Next, a brief comparison between our proposal and previous approaches is made for both OAM-domain and path-domain. Different from other OAM-domain approaches, the cut-off effects from infinite OAM-encoding basis to finite subset is largely alleviated since ancillary guard bands are employed. During the unitary operation, part of the input energy would enter the guard bands, but such bands could be managed to vanish eventually through interference according to meticulously design. The number of required spectral shaper layers grows in $O(N)$ scale according to dimensionality $N$ of unitary operation. The reason is that each spectral shaper could provide $O(N)$ modulation on all the channels in parallel, thus the whole setup can fulfill $N^2$ independent real numbers required by a unitary operation. Compared with the MPLC approach[21,22,48], our design has similar system arrangement of $O(N)$ layers as dimensionality $N$ grows. However, near-unity transformation efficiency and near-perfect fidelity can be achieved theoretically at the same time with our proposal according to the results of numerical simulations. Besides, our proposed scheme can perform parallel unitary operations on different OAM encoding channels simultaneously without increasing system arrangement. It should be mentioned that although our method is based on the mathematic decomposition of the Fourier/diagonal matrix, there is no DFT matrix required. Specifically, our proposed unitary operation scheme for OAM-domain is performed within OAM-domain and angle-domain alternately, while there is no actual DFT transformation between them. Actually, the OAM shaper is employed to perform the diagonal operation, while the DFT matrix is required for the MPLC method. However, since there is no requirement of DFT transformation, our scheme have to be performed within two optical DOF linked by Fourier transformation, *e.g.*, the OAM-domain vs. angle-domain and path-domain vs. wavevector-domain.

For path-domain operation, both of our protocol and Reck's scheme can be adopted and keep similar features of programmability, scalability, and near-unity success probability in theory, although they are based on different matrix decomposition. In this work, our experiment is based on three SLMs to perform the transformation. In our setup, the insertion loss of utilized three SLMs is about 11.9dB. Thus, the scale of our experimental scheme would suffer from the insertion loss of the SLM since more SLMs are required for transformation with higher dimensionalities. It should be mentioned that our proposal is also manageable in both free space optics and integrated circuits. The SLMs could be replaced by three-dimensional set of metasurfaces[49] to achieve a compact and low-loss module for unitary operations. Due to the similar alignment of path channels in our proposal and Reck's scheme, an on-chip version of our design could also be implemented. The key function of Fourier transform could be achieved by the structures of Rowland circle according to the recently reported work[50]. Note that in this design no MMI couplers[29] nor fully connective $N \times N$ blocks[30] are needed. Our proposal would provide an alternative choice in addition to the reported integrated architectures[29,30].

In conclusion, we have proposed and demonstrated a scheme for unitary operation in OAM and optical path domains. A three-layer setup is implemented to evaluate the performance. Field simulation of OAM-domain Hadamard matrices and experimental implementation of randomly generated path-domain unitary matrices have been carried out. For path domain, the fidelity value of $0.97 \pm 0.02$ and transformation efficiency of $0.96 \pm 0.03$ are experimentally evaluated simultaneously.

## Acknowledgments

This work was supported by the National Key Research and Development Program of China (2018YFB2200402 and 2017YFA0303700), the National Natural Science Foundation of China (Grant No. 61875101). This work was also supported by Beijing academy of quantum information science, Beijing National Research Center for Information Science and Technology (BNRist), Beijing Innovation Center for Future Chip, and Tsinghua University Initiative Scientific Research Program. The authors would like to thank Dr. Peng Zhao for valuable discussions and helpful comments.

## ORCID iD

Xue Feng https://orcid.org/0000-0002-9057-1549